\input harvmac.tex
\input epsf

\ifx\epsfbox\UnDeFiNeD\message{(NO epsf.tex, FIGURES WILL BE IGNORED)}
\def\figin#1{\vskip2in}
\else\message{(FIGURES WILL BE INCLUDED)}\def\figin#1{#1}\fi
\def\tfig#1{{\xdef#1{Fig.\thinspace\the\figno}}
Fig.\thinspace\the\figno \global\advance\figno by1}
%
  


\def\CF {{\cal F}}

\def\CN {{\cal N}}
\def\CO {{\cal O}}

\def\CZ {{\cal Z}}
 
 \def\R{\relax{\rm I\kern-.18em R}}
\font\cmss=cmss10 \font\cmsss=cmss10 at 7pt
\def\Z{\relax\ifmmode\mathchoice
{\hbox{\cmss Z\kern-.4em Z}}{\hbox{\cmss Z\kern-.4em Z}}
{\lower.9pt\hbox{\cmsss Z\kern-.4em Z}}
{\lower1.2pt\hbox{\cmsss Z\kern-.4em Z}}\else{\cmss Z\kern-.4em Z}\fi}

\font\cmss=cmss10 \font\cmsss=cmss10 at 7pt


\def\IR{\relax{\rm I\kern-.18em R}}
\def\IZ{\relax\ifmmode\mathchoice
{\hbox{\cmss Z\kern-.4em Z}}{\hbox{\cmss Z\kern-.4em Z}}
{\lower.9pt\hbox{\cmsss Z\kern-.4em Z}}
{\lower1.2pt\hbox{\cmsss Z\kern-.4em Z}}\else{\cmss
Z\kern-.4em
Z}\fi}

\def\b{\beta}

\def\pl{{\it  Phys. Lett. }}

\def\mpl{{\it Mod. Phys.   Lett. }}
\def\np{{\it Nucl. Phys. }}

\def\cmp{{\it Comm. Math. Phys.  }}
\def\r{{\rm Re}}
\def\i{{\rm Im}}

\def\hf{{1\over 2}}
\def\t{\theta}
 
\def\sn{{\rm sn} }
\def\cn{{\rm cn} }
\def\dn{{\rm dn} }
\def\z{\zeta }
 
\def\uy{u_{\infty}}

 \def\l{\lambda}
 \def\11{1\!\! 1}
 

\lref\Kon{I. Kostov, \mpl A 4 (1989) 217.}

 \lref\bax{R. Baxter, Exactly Solved
 Models in Statistical Mechanics, Academic Press, 1982.}

 \lref\bipz{E. Br\'ezin, C. Itzykson, G. Parisi and J.-B.
Zuber, {\it Comm. Math. Phys.} 59 (1978) 35.}

  \lref\DG{
 P. Di Francesco and E. Guitter, {\it Europhys. Lett. }  26 ,
   455 (1994),  {\it  Phys. Rev. } E {\bf 50}, 4418 (1994).}
 \lref\DGK{P. Di Francesco, E. Guitter and C. Kristjansen, 
 Nucl.Phys. B549 (1999) 657.}
  
   \lref\Hoppe{J. Goldstone, unpublished; 
J.~Hoppe,  "Quantum theory of a massless
relativistic surface ...", MIT PhD Thesis 1982, and  Elementary 
Particle Research Journal
(Kyoto) 80 (1989).} 
\lref\FS{{\it B.A.Fuchs,B.V.Shabat,
Functions of a complex variable},
Jawahar Nagar, Delhi; Hindustan Publ.Corp. 1966
}
\lref\BF{P.  Byrd and M.  Friedman,  {\it Handbook of Elliptic Integrals for 
Engineers and Physicists},  Springer-Verlag, 1954.}
 \lref\AS{M. Abramovitz and I. Stegun,
{\it Handbook of Mathematical Functions}, US Dept of Commerce, 1964.}

\lref\KKN{V. Kazakov, I. Kostov  and N. Nekrasov,
\np B557 (1999) 413-442.}
 
 \lref\Naka{ N. Nakanishi, \cmp 32 (1973) 167.}
\lref\EK{B.Eynard and C. Kristjansen, 
{\it Nucl. Phys.} B516 (1998) 529,  cond-mat/9710199.}
 \lref\CMM{ G. Cicuta, L. Molinari and E. Montaldi, \pl B 306 (1993) 245.}
 
\lref\KKinv{V. Kazakov and I. Kostov, 
``Dynamics of vortices in the C=1 string theory",in preparation.}
\lref\hoppe{J. Hoppe, ``Quantum Theory of a Massless Relativistic 
Surface $\ldots$'',  {\it Elementary Particle Research Journal} 
(Kyoto) 80 (1989).}
\lref\FG{P. Di Francesco and E. Guitter, {\it Europhys. Lett.}
 26 (1994) 455.}
\lref\sixvertex{P. Zinn-Justin, cond-mat/9903385; I.K. Kostov,
hep-th/9911023.}
\Title{ SPhT/t00/074  }
{\vbox{\centerline
 {   Exact  Solution of the Three-color Problem }
\centerline{  on a Random Lattice  }
 \vskip2pt
}}
%
\centerline{ Ivan K. Kostov \footnote{$^{\ast}$}{{\tt kostov@spht.saclay.cea.fr}}
\footnote{$^\dagger$}{member of 
CNRS}}
\centerline{{\it C.E.A. - Saclay, Service de Physique 
Th{\'e}orique }}
 \centerline{{\it 
  F-91191 Gif-Sur-Yvette, France}}

 \vskip 1cm
\baselineskip8pt{
 
\vskip .2in
 
\baselineskip8pt{
\noindent
 We  present  the exact solution of the Baxter's three-color problem
 on a random planar graph, using  its  formulation  in
 terms of three coupled random matrices. We find that the number of 
 three-colorings of an infinite random graph is
 0.9843 per vertex.
}

  \bigskip
\leftline{PACS codes: 05.20.y, 04.60.Nc}

\leftline{Keywords: Three-color problem, loop gas, 
random lattices, 2D quantum gravity }

\Date{May, 2000}  

\baselineskip=14pt plus 2pt minus 2pt

\newsec{Introduction }

The Baxter's three-color problem on a regular 
hexagonal lattice \ref\Bax{R. Baxter, {\it J. Math. Phys.}, 
11 (1970) 784, {\it J. Phys.}  A19 Math. Gen. (1986) 2821.}
is one of the classical examples of solvable lattice models.
 The free energy of this model is equal to the 
 number of coloring the links of the lattice with three different colors,
 $A$, $B$ and $C$, so that no links that meet at a vertex carry
  the same color.   Alternatively one can speak of the number of
   three-coloring of the links of the regular triangulated lattice
    in a way that 
    the three sides of every triangle have different colors.
     The problem   is equivalent to a special version of the $O(2)$ 
model, whose partition function is given by  a gas of fully packed 
loops  on the hexagonal lattice, having two different flavors.  
 Baxter also showed that this  
model   solves the problem of coloring the faces of the hexagonal
lattice with four different colors, so that the adjacent faces have 
different colors.
   More recently yet another
geometrical 
interpretation of this model   has been  found, 
namely as the problem  of counting the different foldings of a regular 
triangular lattice \DG.
 
The three-coloring problem can be formulated also for a 
random   3-coordinated planar graph 
(or a random triangulation, in the dual language) \Naka . 
Here one has the freedom to chose different ensembles of planar graphs.
 The  generating function 
for all possible  three-colored planar graphs is given by the  
  three-matrix integral  \CMM 
 \eqn\defZN{\CZ_{N}(\b) =\int_{N\times N} dA \ dB \ dC\ 
 \exp\left\{-N\Tr\left( \hf(A^{2}+B^{2} +C^{2} )-g 
 [ABC+BAC]\right)\right\} } 
  where $A, B$ and $C$ are hermitian $N\times N$ matrix variables.
   
The perturbative  expansion of the free energy  
\eqn\unNF{\CF(g ) \equiv {1\over N^{2} }\log \CZ(g)
= \sum_{  h=0}^{\infty}  
 \sum_{m=1}^{\infty}  \CN_{h} (2m) \ N^{-2h} g ^{2m} }
has the following combinatorial meaning:
$ \CN_{h} (2m)$ is the number of the connected three-colored 
triangulations of a surface of genus $h$, containing $2m$ triangles.
 (The configurations  with nontrivial symmetry group are 
   taken with the corresponding symmetry factors.)
 
 The  model  has still the interpretation as a    
 fully packed $O(2)$ loop gas model, with 
 the restriction that only loops of even length are allowed.
 It can be also interpreted as the four-coloring problem of the 
 faces of a random 3-coordinated graph. 
 The interpretation 
 in terms of foldings does not generalize to this ensemble: it 
  holds only for   planar graphs  whose faces have even number of 
  sides.   
 
  An iterative procedure for evaluating the  coefficients in the
  expansion \unNF\
  has been proposed by B. Eynard and C. Kristjansen \EK.
   They considered, more generally,  
     the $O(2n)$ loop gas model with only even loops, which reduces 
     to the three-color problem  when $n=1$.
     This $O(2n)$ model  is actually identical to  fully packed $O(n)$ loop
     gas model defined   on a 4-coordinated planar graph\foot{More 
     precisely, each 
     4-vertex is visited either by one loop going straight, or by two 
     loops turning at right angle.}.
      It is therefore expected that the  matrix model \defZN\ is 
       in the class of
 universality of the dense $O(1)$ model, 
 which  means that $\CN_h(2m)$   have the  same
 large-$m$ asymptotics   as the 
 number of the non-colored planar graphs:
 \eqn\asymPN{\CN_h(2m) \sim {g_*}^{-2m} \ m^{ {3\over 2} (h -1)-1 },}
  where $g_*$ is the critical value of the coupling $g$.
   The iterative procedure was carried out up to the order 12, 
   with the result
   \eqn\frEn{\lim_{N\to\infty} \CF(g) =2 {g^2/ 2}  +
14   {g^4/ 4\ }+ 138 {g^6/ 6 }+
1608   {g^8/ 8}+20736 { g ^{10}/10 }
   +286452{g^{12} / 12}+...}

 Soon after the work of B. Eynard and C. Kristjansen, the
 exact solution of the matrix model      \unNF\ has been  found
   for  {\it  purely imaginary} coupling $g$ \KKN.  
  An explicit expression for the free energy of the model was
   obtained in terms of elliptic functions, and it was checked that 
   the coefficients in the expansion \frEn\ are correctly reproduced
   \ref\BE{B. Eynard, private communication.}
   However, it was not  possible to
    to carry out the   analytic continuation of this solution to
   real values of $g$.

In this paper we present the direct  solution of the   matrix 
model \unNF\    for real coupling $g$, using a method similar to the 
one applied in \KKN. 
We find    the expected critical behavior of a $c=0$ matter 
coupled to 2D gravity and 
 evaluate the critical coupling $g_{*}$.
 Finally we we explain why the cases of real and imaginary 
 coupling $g$ are not related by analytic continuation.

  \newsec{Saddle point equations for the matrix integral}

After integrating over the $B$ and $C$ matrices,  the partition 
function reduces to the following  integral
  over the eigenvalues $\l_{1} , \ldots, \l_{N} $ of the 
 matrix $A$:

\eqn\twpa{ \CZ_{N} ( \b)=  \b^{N^{2} } 
\int \prod_{k=1}^N {d\lambda   _k \ e^{-\hf N\lambda_{k} ^{2} }\over 
g^{-2} -4\lambda_{k}^{2}   } \ 
 \ \prod_{i< j}  {( \lambda   _i - \lambda   _j)^{2} 
\over  g^{-2} -( \lambda   _i + \lambda   _j  )^{2} } .}
The saddle-point spectral density
 \eqn\RhO{\rho(\lambda    ) =\lim_{{N\to\infty}} 
   \left\langle  {1\over N} \sum_{i=1}^{N}\delta(\lambda   -\lambda   
   _{i})\right\rangle }
  is an  even function  supported by a symmetric 
interval $[-\Delta,\Delta]$, where $\Delta $ is a function of 
the coupling $g$, 
such that $\Delta|_{g=0}=2$. At $g=0$ it is given by the Wigner's 
semi-circle law: $\rho(\l)|_{g=0} = {1\over 2\pi} \sqrt{4-\l^{2}}$. 
It is completely determined by 
 the saddle-point 
 equation
\eqn\sdlptt{
  \lambda     = \int_{- \Delta }^{\Delta }
\!  \! \! \! \! \! \! \! \! \! \! \! -   d\lambda ' \ 
  \rho(\lambda   ')\ \left[ {2\over \lambda   -\lambda   ' } 
 -    {1\over  \lambda   +\lambda   ' +g^{-1}  }
-  {1\over \lambda    +\lambda   '-g^{-1}  }
   \right]
} 
and  the normalization condition 
$\int d\l \ \rho(\l) =1.$
Introducing the analytic function
\eqn\REZ{   W( z) =   \int_{ -\Delta }^{\Delta} {d\lambda     \ 
  \rho(\lambda    ) \over z - \lambda  } 
 }
  we write the   saddle point equation  as a functional condition
  for  $W(z)$:
\eqn\cauSat{   W(\lambda    +i0)+  W(\lambda    -i0) +  
W(-1/g - \lambda    )
  +   W(1/g -\lambda     )   
 =   \lambda     ,  \qquad   \lambda    \in [-\Delta, \Delta] . }
The density  $\rho(\l)$  is an even  function of $\l$, which means 
that    \REZ\ is 
antisymmetric:
 \eqn\antisW{W(z) = -W(-z). }
 The saddle point equation can be therefore   rewritten as
\eqn\cauS{   W(\lambda    +i0)+  W(\lambda    -i0) -
W(  \lambda  +1/g  )
 -   W( \lambda -1/g    )   
 =   \lambda   ,  \qquad   \lambda    \in [-\Delta, \Delta] .}
The  
 normalization condition for the density 
 fixes the first coefficient in the expansion of $W(z)$ at infinity:  
\eqn\normden{ W(z) = {1\over z} + 
{\left\langle {\Tr\over N} A^{2} \right\rangle\over z^3} + \CO(z^5) .}

As mentioned in \EK , knowing the  the quantity
 $ \langle {\Tr\over N} A^{2} \rangle$
is enough to solve the 
three-coloring problem, because   it is related to the free energy 
by
\eqn\AdveF{ \left\langle {\Tr\over N} A^{2} \right\rangle = 1+g{d\over 
dg } \CF(\b ).
 }

 \newsec{Solution of the saddle point equation }

The   four-term functional equation \cauS\ with the boundary condition 
 $W(z)\sim 1/z$ at infinity is 
sufficient to determine $W(z)$. Introducing the   function  
  \eqn\Jofu{ \eqalign{\z(z) 
 &= 2 g^{-1} [ W(z+ 1  /2g) -W(z - 1/2g) ]  +  z^{2}  ,\cr
 } }
 we write  \cauS\ as  a two-term  difference equation
 \eqn\fueqJ{  \z(\lambda    +  1/2g\ \pm i0) =   \z(\lambda   -1/2g\ \mp i0),
   \qquad   \lambda    \in [-\Delta, \Delta] .  }  
 The function $\z(z)$ 
 is  real and symmetric: 
   \eqn\realityetc{\z(z ) =\z(-z )  =  \overline{\z(\bar z}),}
 possesses two cuts along the intervals
 $[ -\Delta-{1\over 2g} , \ \Delta- {1\over 2g}]$ and 
  $ [-\Delta+{1\over 2g}, \
\Delta+{1\over 2g}]$ on
 the real axis, and 
expands  at $ z\to +\infty$  as 
\eqn\Jinfty{ \eqalign{\z(z) &= z^{2}   \ - \ {2  \over 
g^{2} z^{2} }\ -\ 
  { ({ 1 \over  2g^{2} } + {6  }W_{2}) 
\over g \ z^{4}} \ +\ \CO(z^{-6} ).\cr}}

  From the  symmetry of   $\z(z)$ it follows  that 
it  is  real when $z$ is real and outside the cuts, and also when $z$ 
is purely imaginary.  
Therefore the map  $z\to \zeta  $ 
 transforms the quadrant $  \{ \i z> 0, \ \r z > 0\}$ into the 
  to the upper  semi-plane $  \{ \r\zeta>0\}$, cut along  an arc   
  connecting the points $c$ and $b$,  
  where $b$ is  some complex number (Fig. 1). At the endpoint $c$, 
  the arc should 
  be perpendicular to the real axis.
We denote the images of the  special points of the map 
\eqn\zspeci{z_{1}=0, \  \ z_{2 } = {1\over 2g} - \Delta, 
\ \ z_{3}=  {1\over 2g}, \  \
z_{4} = {1\over 2g} + \Delta, }
 by
  \eqn\zetaspeci{ a={\z(z_{1})} , \  \ b=\z(z_{3}), \ \ 
  c=\z(z_{2})=\z(z_{4}).   }
  
%
\vskip 20pt
\hskip 20pt
\epsfbox{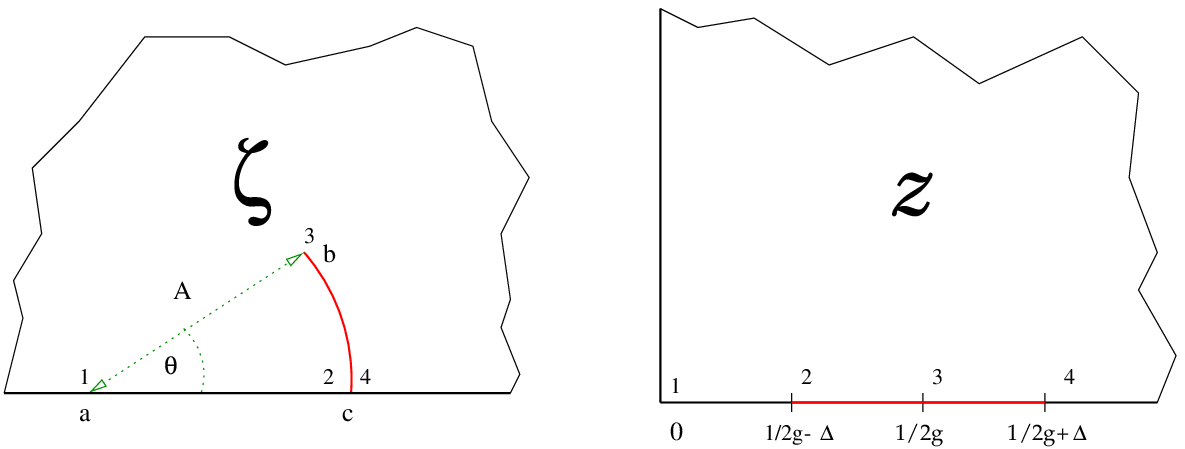}
\vskip 5pt

  \centerline{Fig.1:    The map $z\leftrightarrow \zeta$.}

\vskip 20pt

  It is easy to see that the inverse function $z(\z)$ can be
 obtained 
  as the integral\foot{Here we are   following the argument by J. Hoppe
  in \hoppe.}
   \eqn\integrl{z=\hf \int_{a}^{\z}{dt \ (t-c)\over
   \sqrt{(t-a)(t-b) (t-\bar b)}}.}
  The asymptotics \Jinfty\ is satisfied if
 \eqn\candc{    a+b+\bar b -2c=0, \ \ \ 2c^{2}-a^{2}-b^{2}-\bar b^{2}= 
 6/g^{2}     .}

   The integral \integrl\ is a standard elliptic integral. Denoting  
    \eqn\zedUU{ \cn u \equiv \cn (u, k) = {A-\z +a\over A+\z-a}, 
    \ \ \ b -a = Ae^{i\t}, \ \ \ k=\cos(\t/ 2).}
   we have ( \BF ,  239.00, 239.07 and 341.53)
  \eqn\zedU{\eqalign{-{z\over \sqrt{A}} &= \ \  { a- c- A\over 2 A }\ u 
   +    \quad    \int_{0} ^{u} {du \over 1+\cn u} \cr
   &= C u - 
   {H'(u,k)\over H(u,k)} + {\dn (u,k) \over \sn (u,k) }  \cr}
  }
   where
   \eqn\CcCC{ C=  {A+a-c\over 2A} -{E\over K} .}
The elliptic parameters $u_{1}, \ldots, u_{4}$  of the special points
 of the map $z\to \z$ 
are (see Fig. 2) 
 \eqn\uspeci{u_{1}=0, \ u_{2}\in[0,K],\ u_{3}  = K+iK',  u_{4}= 2 u_3 - u_2,}
 and the  point  $z\to\infty$   $(\z\to\infty)$   corresponds to  
 $\uy = 2iK'$. 

  \vskip 20pt
  \hskip 100pt
  \epsfbox{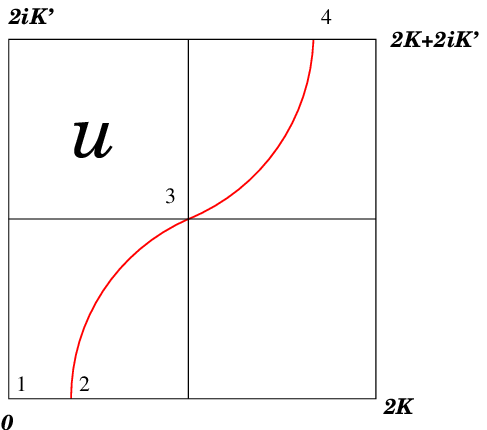}
  \vskip 5pt

  \centerline{Fig.2:    The domain of the  $u$-variable.}  
\vskip 20pt
 \noindent From \zspeci\ we determine the  constants $A$ and $C$:    
 \eqn\zbeta{z_3= z(K+iK')  = \sqrt{A}\left[-CK+i\left({\pi\over 2K}+ CK'\right)
 \right] = {1\over 2g},}
  which gives
 \eqn\CcAa{ C=- {\pi\over 2KK'} ,  \ \ \sqrt{A} =  {\b\over \pi} K'.}
 Then,
 using the Legendre's relation $ EK'+E'K-KK'=\pi/2$, we write \CcCC\ as
\eqn\CaA{{c-a\over A} = {2E'-K'\over K'}.}
\def\tz{\tilde z}
\def\ttz{\tilde \z}

 The final formula is 
\eqn\ZedZt{ \eqalign{\tz (u) & \equiv\ \ \  {z(u) \over 
\sqrt{A}}\ \ \ =\ \  \
{\pi \over 2KK'} u +  {H'(u,k)\over H(u,k)}-
{\dn (u,k) \over \sn (u,k) }
    \cr
 \ttz(u) &\equiv    {\z(u)-a \over A}=   
   \ {1-\cn (u, k) \over 1+\cn (u, k)}.\cr}}\bigskip
  Note the very useful relation  
   \eqn\zprime{ 2{d\tz\over d u} = \ttz(u) 
   -{2E'-K'\over K'}.
   }
  
   Now it remains to determine the elliptic modulus as a function of 
   $g$. It is fixed by the ${1\over \tz^2}$-term of the  
   expansion  
 \eqn\asuy{\ttz(\tilde z)= 
 \tz^{2 }  -{a\over A}
 - {2 \over g^2 A^2} {1\over \tz^2} 
 -\  {  {1 \over 2g^{2}} 
 +{6} \left\langle {\Tr\over N} A^{2} 
 \right\rangle \over g^{2} A^3} {1\over \tz^4} 
   +\
   ...}
  
 Comparing the Laurent expansions of   $\ttz(u)$ and $\tz(u)$ at the point $\uy =
  2iK'$ (the expansion of $z(u)$ is most easily obtained using 
  the relation \asuy), we find
   \eqn\modluS{  {g^{2}\over 2 } \ \left({\pi\over K'}\right) ^{4}   =
\left(   {2E'-K'\over 2K'} +{1-2k^{2}\over 3}  \right)^{2}
-{ 1 -16 k^2   k'^2   \over 36} .}
The expansion \unNF\ of the free energy   can be 
obtained using \AdveF\ from the ${1\over \tilde z^{4}}$-term  of the expansion 
\asuy.
The Gaussian limit $g\to 0$ corresponds to 
   $ k'={4g} + {g  ^{2}}+\ldots \to 0$. Therefore,  
   it is   convenient first to 
express the solution \ZedZt\ in terms of the dual modulus. 
Using the duality relations
$$ H(u,k) = -i  \sqrt{K/ K'} e^{-\pi u^{2}/4KK' }    H(iu, k')  $$
$$\cn(u,k) = {1\over \cn(iu, k')}, \ \ \ {\rm ds} (u,k) = i  {\rm ds}(iu, k')$$
we write 
\eqn\Dzedzt{\eqalign{
 \tz (u)  &= 
 i \left[{H'(iu,k')\over H(iu,k')}-
{\dn (iu,k') \over \sn (iu,k') }\right],
    \cr
 \ttz(u)  &=   
   \ { \cn (iu, k')-1 \over  \cn (iu, k')+1}.\cr}}
\bigskip\bigskip
The  large-$N$ limit of the  free energy $\CF$  is obtained 
from the asymptotics \asuy. This is purely technical exercise 
and we will not do it here.
 Instead, we will obtain the asymptotical behavior of the 
free energy when the volume of the graph diverges.
  
\newsec{Critical behavior}

  The Taylor expansion of the free energy  in $g$ is convergent up 
  to the critical point $g_{*}$, where the nearest singularity of
  $\CF(g)$ is located.  Since the free energy is an analytic function 
  of $k^{2}= 1-{k'_{*}}^{2}$, the  critical point is determined by $d \b/dk =0$. 
   The function $\b(k')$ has a minimum at ${k'_{*}}^{2}= 0.826114
$ where 
 \eqn\crP{2E(k'_{*}) =K(k'_{*})    
 .}
  In the vicinity of the critical point ($k'=  k'_{*}+ \delta 
k'$) 
 \eqn\critN{ {g^{2}  \over g_{*}^{2}} =   1- 
{1-2(k'_{*}  k_{*})^{2}\over ( k_{*} k'_{*})^{2} }\delta 
k'^{2} + \CO( \delta 
k'^{4} ), }
where 
\eqn\gstar{ 
 g_{*}   =  { K ^2(k'_{*})  \over    \pi ^2  \sqrt{6} }\   = 
  \  {\left( K^{2}\right) _{2E=K}  \over\pi ^{2} \sqrt{6} }.
}
  By \CaA\ this is exactly the point  where the  the right end of the 
  left cut  of $\z(z)$  
  touches the left end of the right cut.  
  At this point   $u^{*} _{2} =u^{*} _{1} =0$, $c^{*} =a^{*} $ and 
  $\Delta ^{*} = {1\over 2g_*}$.
   Finally, it is evident that the critical singularity of the free energy 
 is $\CF(g)- \CF(g_{*})  \sim ( g_{*}-g)^{{5/2}}$, which implies
 \asymPN.

\newsec{Conclusion}
 
We have found that the coefficients of the free energy \unNF\   
 grow as 
 $$ \CN_{0} (2m)  \sim g_{*}^{-2m} \  m^{-7/2}.$$
  On the other hand
 it is known \bipz\ that the number of 3-coordinated 
 planar graph with $2m$ vertices grows as
 $  (12\sqrt{3})^{n} \ n^{-7/2}$.
 Therefore the number of  three-colorings per vertex of an
 infinite random  3-coordinated planar graph is
  equal  to
 \eqn\tricolR{{1/g_*\over \sqrt{12\sqrt{3}}}=0.984318\ldots
 .}
 For a large but finite planar graph, the  
 number of three-colorings grows linearly with  the number of vertices 
 $A=2m$, up to a term $\CO({1\over A})$. There is no logarithmic 
 corrections. 
 The fact that the number of three-colorings is slightly less than 
 one is explained by the fact that not all 3-coordinated graphs are 
 three-colorable.
 
 In terms of the $O(2)$ loop gas, eq. \tricolR\ gives the entropy per vertex
 of the 
 gas fully packed loops on a random 3-coordinated graph,  having
 two different flavors and even length.
 This is to be compared with the  entropy of the gas of fully packed loops
 with $n$ flavors and no restriction on the length  \Kon
 $${1/g_*^{_{^{ O(n)} } } \over \sqrt{12\sqrt{3}}}= 
 {2\sqrt{2(2+n)}\over \sqrt{12\sqrt{3}},}
 $$ 
   which is equal to  $1.24081$  for $n=2$.

   In order to interpret our result in term of the   $O(1)$ loop gas
   on a 4-coordinated planar graph, we recall that the number of 
   planar 4-coordinated graphs with 
   $m$ vertices grows as $12^{m }m^{-7/2}$ \bipz. Therefore the
   entropy per vertex of the fully packed 
    non-oriented loops on a 
   4-coordinated
   random graph is  is given 
   by
   $$ {1/g_{*} ^{2} \over 12} = 1.62593\ldots.$$

  Finally, let us mention that   the integral \twpa\
  is related to  the Baxter's 6-vertex model on a random lattice, 
  which has been recently solved exactly \sixvertex.
  In the parameterization 
  $$a=b=1, \ c=-2\cos\b$$
   of the vertex  weights, 
  the integral \twpa\ corresponds to the limit
  $\b\to 0$, where $\b$ is assumed to be purely imaginary.
  The parameter
  $$\Delta = {a_2+b^2-c^2\over 2ab}= -\cos 2\b$$
   achieves the value $-1$
  from the left, which means that  we are approaching 
   the boundary between the regimes III and IV  from the side 
   of the regime IV. This  regime IV is characterized by
   a finite correlation length,
  which is confirmed by our solution.  On the other hand, the 
  matrix integral considered in \KKN\ can be  identified  with 
  the limit $\b\to 0$   with $\b >0$,   which means that the
    the   boundary III/IV  is achieved from the
    critical regime III characterized by an infinite 
    correlation length. The fact that  the   real and imaginary 
    values of the coupling $g$ are associated with different regimes of the 
    six-vertex model explains why our solution
    cannot be obtained from the 
      solution found in \KKN\   by analytic continuation
      $ig\to g$. Even if the
    perturbative expansions in $g^2$ are identical, they 
    differ by nonperturbative terms.     
      
    \bigskip
    
    \noindent
    Note added: 
    After this manuscript was  to the publisher,  the author 
    learned that Bertrand Eynard and Charlotte Kristjansen 
    found independently the exact solution of the problem.

  \newsec{Acknowledgments}
 
  \noindent  
  
  The author thanks P. Di Francesco for useful discussions. 
  This research
  is supported in part by European  network EUROGRID
  HPRN-CT-1999-00161.


 \listrefs

\bye